\documentclass[10pt]{iopart}

\usepackage{gensymb}
\usepackage{graphicx}
\usepackage{caption}
\usepackage{subcaption}

\usepackage{textcomp}

\begin{document}

\title[Observations of 2D Doppler backscattering on MAST]{Observations of 2D Doppler backscattering on MAST}

\author{D. A. Thomas$^{1,2}$, K. J. Brunner$^{3}$, S. J. Freethy$^{1,2}$, B. K. Huang$^{2,3}$, V. F. Shevchenko$^{2}$ and R. G. L. Vann$^{1}$}

\address{$^{1)}$\textit{York Plasma Institute, Department of Physics, University of York, York \\ YO10 5DD, U.K.}\\ 
               $^{2)}$\textit{CCFE, Culham Science Centre, Abingdon, Oxon, OX14 3DB, U.K.} \\ 
               $^{3)}$\textit{Centre for Advanced Instrumentation (CfAI), Department of Physics, Durham University, Durham, DH1 3LE, U.K.}       }
\ead{dadt500@york.ac.uk}
\vspace{10pt}
\begin{indented}
\item[]June 2015
\end{indented}

\begin{abstract}

The Synthetic Aperture Microwave Imaging (SAMI) diagnostic has conducted proof-of-principle 2D Doppler backscattering (DBS) experiments on MAST. SAMI actively probes the plasma edge using a wide ($\pm$40$\degree$ vertical and horizontal) and tuneable (10-35.5 GHz) beam. The Doppler backscattered signal is digitised in vector form using an array of eight Vivaldi PCB antennas. This allows the receiving array to be focused in any direction within the field of view simultaneously to an angular range of 6-24$\degree$ FWHM at 10-34.5 GHz. This capability is unique to SAMI and is an entirely novel way of conducting DBS experiments. In this paper the feasibility of conducting 2D DBS experiments is explored. Initial measurements of phenomena observed on conventional DBS experiments are presented; such as momentum injection from neutral beams and an abrupt change in power and turbulence velocity  coinciding with the onset of H-mode. In addition, being able to carry out 2D DBS imaging allows a measurement of magnetic pitch angle to be made; preliminary results are presented. Capabilities gained through steering a beam using a phased array and the limitations of this technique are discussed.

\end{abstract}

%
%
%
\maketitle
%
\ioptwocol

\section{Introduction}
\label{sec:introduction}





DBS was developed from conventional reflectometry when fluctuations propagating perpendicular to the magnetic field were causing phase runaway while the antenna was oriented oblique to the cutoff surface \cite{holzhauer1}. DBS maintains many of the advantages of reflectometry including: infrequent access to the machine needed, only a small amount of port space required and high spatial and temporal resolution. As with most microwave diagnostics, DBS experiments can be conducted using antennas constructed from materials that are resistant to high heat and neutron flux environments. In addition, waveguides allow electronic components to be delocalised from the reactor and placed behind neutron shielding if necessary. Therefore, DBS is one of the few plasma diagnostic techniques that is suitable for deployment on next generation fusion devices; this renders its development crucial.

DBS experiments have been used to measure the perpendicular velocity profiles of turbulence structurures and turbulence amplitude on ASDEX Upgrade (AUG) \cite{conway1}, DIII-D \cite{hillesheim1}, W7-AS \cite{hirsch1,hirsch2}, EAST \cite{zhou1}, HL-2A \cite{weiwen1}, LHD \cite{tokuzawa1}, L-2M \cite{pshenichnikov1} and MAST \cite{hillesheim2}. Turbulence velocity profiles are of interest as radial velocity shear has been shown to influence the stability properties of drift-type instabilities, for example \cite{burrell1}. In addition, if the $E \times B$ velocity dominates the turbulence velocity then the radial electric field can be calculated. Mechanically steerable mirrors and antennas have allowed $k$ spectra to be measured in addition to turbulence velocity on DIII-D\cite{schmitz1}, TJ-II \cite{happel1}, Tore Supra \cite{hennequin1} and AUG \cite{happel2}. DBS has also been used to study the toroidal and radial structure of geodesic acoustic modes on DIII-D \cite{hillesheim1,hillesheim3,wang1}, AUG \cite{conway2}, TCV \cite{huang1}, Tore Supra \cite{sabot1} and FT-2 \cite{gurchenko1}. The perpendicular velocity, size and quasi-toroidal mode numbers of filaments in the edge region were determined using DBS on Globus-M \cite{bulanin1}.

\begin{figure}[h!]
\centering
\includegraphics[width=0.5\textwidth]{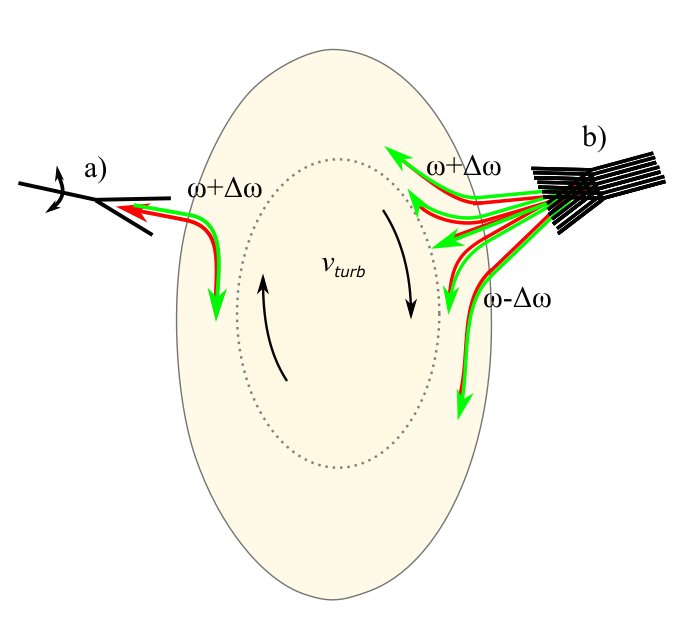}
\caption{Poloidal cross-section cartoon of DBS. The plasma is indicated by the beige region. The incident probing beams are shown in green and the backscattered beams are shown in red. The direction of the turbulence velocity is indicated by the black arrows. The normal incidence O-mode cutoff is indicated by the dotted black line.  a) A steerable conventional single horn DBS experiment. b) The SAMI diagnostic probes the the plasma with a broad beam and receives backscattered radiation from multiple directions and Doppler shifts on eight phase sensitive antennas.}
\label{cartoon}
\end{figure}

A conventional DBS experiment comprises a horn antenna launching a beam oriented perpendicular to the magnetic field and \textit{oblique} to the normal incidence cutoff surface (see Figure~\ref{cartoon}a). The returned signal is Bragg-backscattered off turbulent structures elongated along the magnetic field lines. Due to refraction the backscattering occurs at a layer that is shifted radially outwards from the normal incidence cutoff. Backscattering occurs according to the Bragg condition near the cutoff

\begin{equation}
\label{eq:perp}
K_{\mathrm{\perp}} \simeq -2k
\end{equation}

\noindent
where $K_{\mathrm{\perp}}$ is the \textit{binormal} component of the density perturbations perpendicular to the equilibrium magnetic field and density surface normal. The incident wavenumber of the probing beam \textit{at the scattering location} is given by $k$. As the probing beam propagates through the plasma into regions of higher density $k$ will decrease. Near the cutoff backscattering will occur continuously along the path of the beam. However, the amplitude of the turbulence decreases with wavenumber as $K^{-3}$ or faster \cite{hennequin2} and the scattering efficiency is $\propto K^{-2}$\cite{gusakov1}. In addition the electric field of the probing beam increases as the beam propagates towards higher density \cite{blanco1, gusakov1}. These two effects highly localise backscattering to the region of lowest possible $k$. If the probing frequency is denoted by $\omega$, and Doppler shift by $\Delta \omega$; $\Delta \omega $ is then linearly proportional to the perpendicular velocity of the density turbulence given by $v_{\mathrm{turb}}=v_{E \times B}+v_{\mathrm{phase}}$, where $v_{E \times B}$ is the plasma $E \times B$ velocity and $v_{\mathrm{phase}}$ is the phase velocity of the turbulent structures. In many cases \cite{conway1,hirsch1} the $E \times B$ flow dominates allowing the radial electric field, $E_{\mathrm{r}}$, to be calculated using $E_{r}=-v_{E\times B}B$ \cite{ida1}.

In many conventional DBS experiments the back-scattered radiation is detected by a single horn antenna (Figure~\ref{cartoon}a). In the linear regime the scattered power is proportional to the density fluctuation power \cite{gusakov2} and the beam is Doppler shifted by the lab frame propagation velocity of the turbulent structures. In order to change viewing orientation, and therefore scattering wavenumber, a narrow beam is mechanically steered between shots \cite{hillesheim2}. In contrast, the Synthetic Aperture Microwave Imaging Diagnostic (SAMI) conducts DBS experiments by launching a broad ($\pm40\degree$ horizontal and vertical) beam containing both O and X-mode polarisations using an antipodal Vivaldi PCB antenna \cite{gibson1,langley1} and receives the backscattered signal on eight antenna channels simultaneously using an array of 8 Vivaldi antennas (Figure~\ref{cartoon}b). The signals received on each of the antennas are then split into in-phase, $I=A \cos \phi$, and quadrature, $Q=A \sin \phi$ components (where $A$ and $\phi$ are the amplitude and phase of the scattered electric field respectively). The I and Q components for each antenna channel are then digitised. Digitising phase and amplitude from eight antennas simultaneously allows SAMI to focus the receiving beam in any direction within $\pm40\degree$ post shot: this allows SAMI to look in all directions within this range simultaneously. This capability is unique to SAMI and is an entirely novel way of conducting DBS experiments. For Doppler backscattering to occur the probing beam must be aligned perpendicular to the magnetic field. On spherical tokamaks, due to the large magnetic pitch angle, optimal alignment of the beam for backscattering is a major concern during conventional DBS experiments. Being a 2D DBS device, not only does SAMI forgo this problem, using the orientation of the backscattered signals allows SAMI to measure the magnetic pitch angle as will be discussed further in Section~\ref{sec:pitch}. Despite these many advantages, phased array systems are subject to many limitations which will be discussed in Sections~\ref{sec:sumsig}, \ref{sec:results}, \ref{sec:conclusion} and \ref{sec:furtherwork}.

\section{Method}
\label{sec:sumsig}

\subsection{The SAMI diagnostic}

\begin{figure}[h!]
\centering
\includegraphics[width=0.507\textwidth]{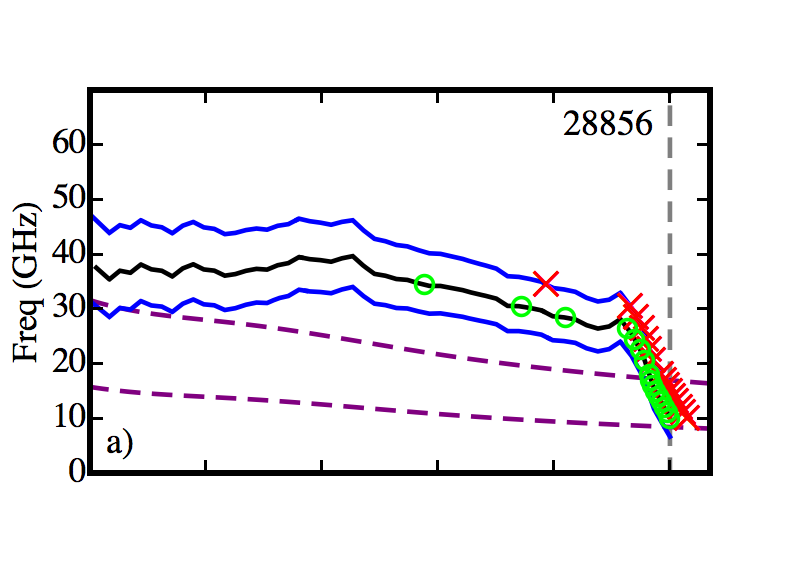}
\\
\vspace{-18pt}
\includegraphics[width=0.5\textwidth]{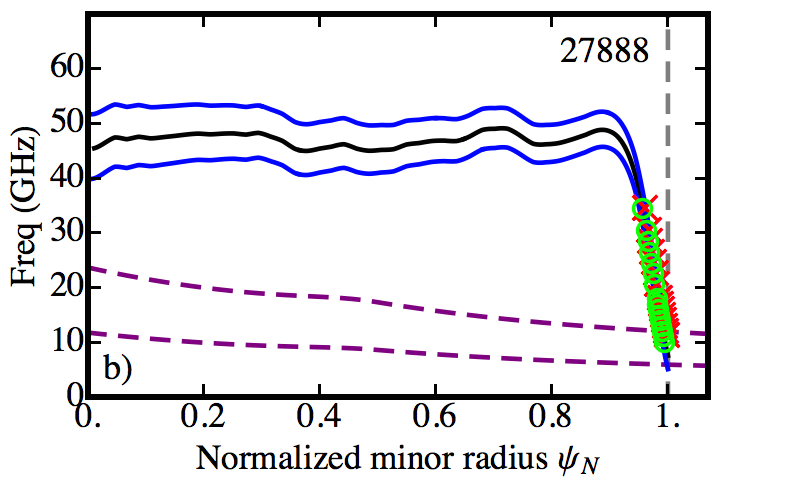}
\vspace{-18pt}
\caption{Normal incidence cutoffs for two MAST shots: a) 28856 in L-mode and b) 27888 in H-mode. Blue lines indicate the X-mode upper and lower density cutoffs. The black lines indicate the plasma frequency cutoff. The dashed purple lines indicate the first and second cyclotron harmonics. The dashed grey line indicates the position of the LCFS. The green circles indicate the positions of the normal incidence O-mode cutoffs. The red crosses indicate the locations of the normal incidence X-mode cutoffs. The density profile data is from the MAST 130 point Thomson Scattering (TS) system \cite{scannell1}. \label{fig:reflPntPlot} }.
\end{figure}

\noindent
In addition to SAMI, two DBS experiments have been conducted on spherical tokamaks: one focusing on the edge \cite{bulanin1}, one focusing on the core \cite{hillesheim2}. Both of these experiments steered their probing beam mechanically. SAMI acquires data at one RF frequency at a time  and can switch between frequency channels during the shot in any order (16 frequencies available between 10 and 35.5 GHz) primarily probing the edge region of the MAST plasma (see Figure~\ref{fig:reflPntPlot}).

\begin{figure}[h!]
\centering
\includegraphics[width=0.4\textwidth]{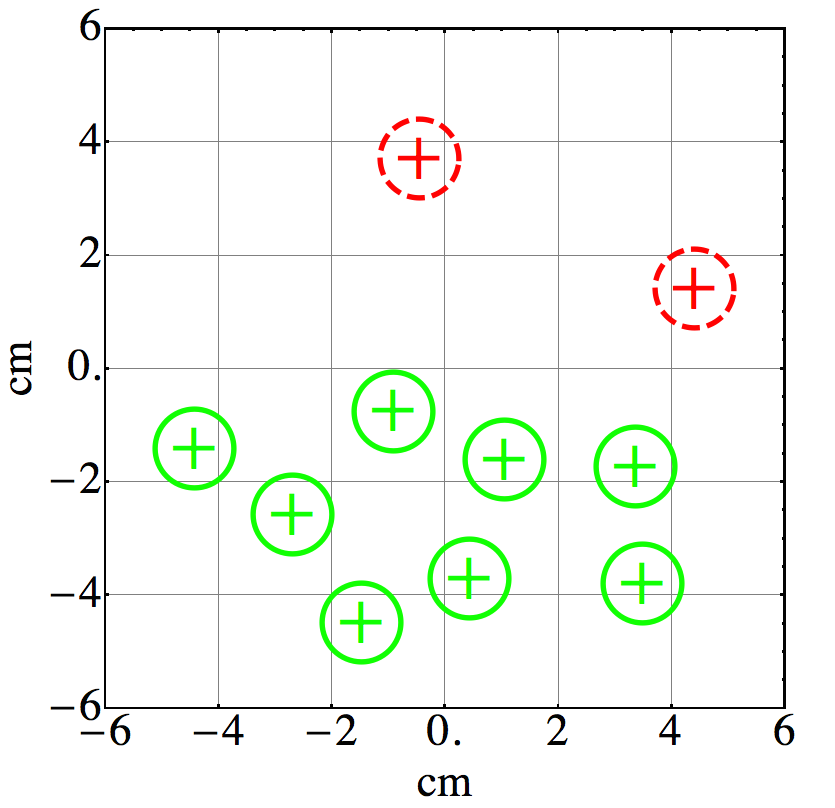}
\caption{The layout of the two SAMI emitting (red dashed crossed circles) and eight receiving antennas (green crossed circles) in the SAMI array plate as viewed from the back of the array.\label{sami_array}}
\end{figure}

SAMI uses an array of independently phased antennas (see Figure~\ref{sami_array}). Two active probing antennas are used; each one launching a different IF probing frequency. The backscattered signal is received by eight antennas simultaneously. Both real and imaginary parts of the electric field are down-converted in frequency by a heterodyne receiver and digitised by a 14 bit 250 Mega samples per second FPGA-controlled digitiser.  SAMI is a wide-field imaging device with a routine field of view of $\pm40\degree$ vertically and horizontally. SAMI probes the plasma with an emitting antenna at either 10 or 12 MHz above the second harmonic local oscillator frequency. The SAMI array is positioned behind a vacuum window made of fused silica. At present, the active probing antennas launch linearly polarised beams which excite X and O-mode radiation in the plasma. The receiving antennas digitise the backscattered signal along a single polarisation; therefore O and X-mode cannot be separated. In Section~\ref{sec:furtherwork} we will discuss the error which this can introduce into our measurements. In future experiments SAMI will be upgraded to duel polarisation antennas which will allow for O and X-mode separation.

\begin{figure}[h!]
\centering
\includegraphics[width=0.5\textwidth]{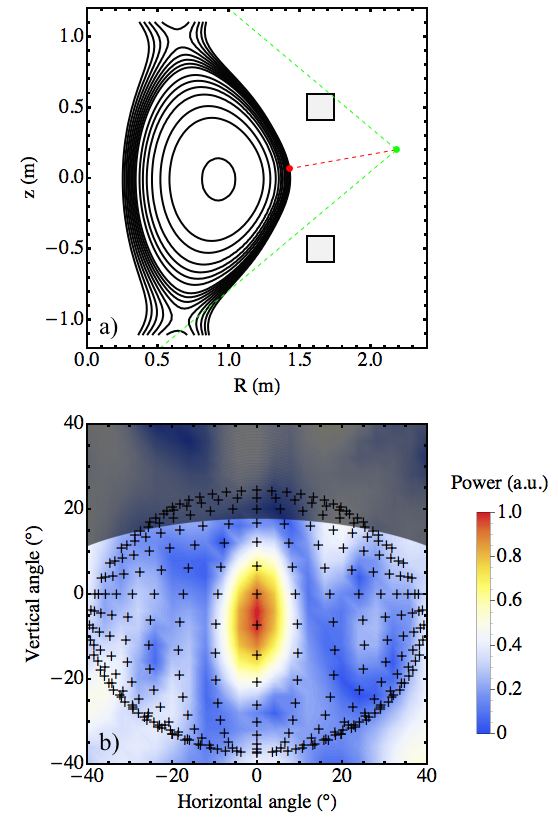}
\vspace{10pt}
\caption{a) Polodial cross-section of the SAMI installation on MAST. Normal incidence O-mode plasma density cutoff surfaces are plotted in black and the upper and lower poloidal field coils are indicated by black squares. The vertical SAMI field of view ($\pm40\degree$) is indicated by the green dashed lines. The position of the SAMI array is marked with a green dot. The position of the test source is shown by a red dot. The test source and the SAMI array are connected by a red dashed line. The normal incidence cutoff surfaces were calculated using data from the MAST TS system and EFIT \cite{lao1,appel1}. b) Normalised linear intensity of the SAMI point spread function at 16 GHz in image coordinates. The position of the normal incidence 16 GHz O-mode cutoff surface is indicated by black crosses. The dark region at the top of the plot shows the field of view which is obscured by the upper poloidal field coil. \label{SAMIinstallation}}
\end{figure}

Figure~\ref{SAMIinstallation}a shows a poloidal cross section of the installation of SAMI on MAST. The entire vertical extent of the plasma is visible to SAMI apart from the top $20\degree$ which is blocked by a poloidal field coil. Figure ~\ref{SAMIinstallation}b shows the response of the system to a point source emitting at 16 GHz. The power received at the SAMI array is recorded as a function of $\pm40\degree$ horizontal and vertical viewing angles. Beam forming has been used to focus the receiving beam at each horizontal and vertical angle in the field of view. Note that although the intensity maximum measured by SAMI is in the correct angular location, the side-lobe power can be up to 50$\%$ of the maximum. This high side-lobe level results from SAMI being limited to eight receiving antennas. Future 2D DBS systems could use a greater number of receiving antennas allowing them better directivity as will be discussed further in Section~\ref{sec:furtherwork}.

SAMI can also operate in passive imaging mode, although this is not discussed in this paper. For SAMI passive imaging experiments we direct the reader to the relevant papers \cite{freethy1,shevchenko1, freethy2}.


\subsection{Beam forming}
\label{sec:beamforming}

The image inversion algorithm employed on the SAMI active probing data is based on the beam forming technique. Beam forming involves applying a phase shift to each of the antenna channels so that when the phase shifted signals are summed together, constructive interference occurs in the direction you wish to focus your beam. The synthesised beam signal, $S^{B}$, can be written as

\begin{equation}
\label{eq:sumsig}
S^{\mathrm{B}}(t,\theta,\phi)=\sum\limits_{i=1}^{N} S_{i}^{\mathrm{A}}[t;\psi_{i}(\theta,\phi)]
\end{equation}

\noindent
where $S_{i}^{\mathrm{A}}$ is the signal from the $i^{\mathrm{th}}$ antenna and $\psi_{i}$ is the phase shift applied to the $i^{\mathrm{th}}$ antenna in order to focus the beam in the chosen horizontal ($\theta$) and vertical ($\phi$) directions. In order to then study the spatial distribution of intensity, $S^{\mathrm{B}}$ is integrated over an exposure time interval $\Delta t$ and the result is squared to generate one pixel in an intensity map, $I$, so: 

\begin{equation}
\label{eq:intmap}
I(\theta,\phi) = \left\{ \int_{\Delta t} S^{\mathrm{B}}(t;\theta,\phi) dt   \right\}^{2}
\end{equation}

\noindent
By doing this over a range of horizontal and vertical viewing angles a 2D map of intensities, such as that shown at the Figure~\ref{SAMIinstallation}b, is calculated. 

\section{Results}
\label{sec:results}

\subsection{Data analysis techniques}
\label{dataanalysistechniques}

Analysis of SAMI active probing data differs from that used on conventional DBS data. Limited to eight receiving antennas only a \textit{partial} suppression of signal from outside the chosen probing direction is possible. This can make the spectra of received signals difficult to interpret.

\begin{figure}[h!]
\centering
\includegraphics[width=0.5\textwidth]{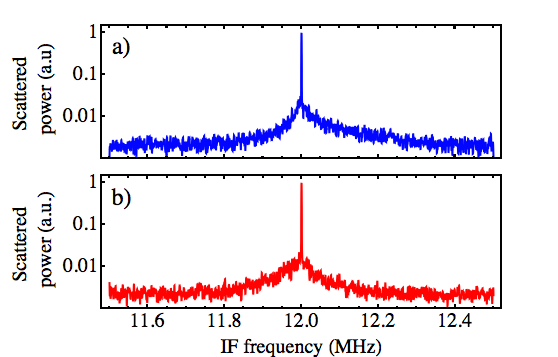}
\caption{
a), b) Spectra of the receiving 16 GHz beam around the active probing frequency focused at the points of most blue and red-shifted power imbalance respectively. Maximum blue and red-shifted power imbalance were observed to be at (-8$\degree$, -12$\degree$) and (0$\degree$, -28$\degree$) in image coordinates respectively. Data taken between 290 and 310 ms into MAST shot 27969. 
\label{fig:spectra}}
\end{figure}

Figures~\ref{fig:spectra}a and \ref{fig:spectra}b show spectra at 300 ms into MAST shot 27969 when the 16 GHz beam was focused at the points were the values of blue minus red and red minus blue-shifted power where at their greatest respectively. The red and blue extrema were observed at (-8$\degree$,-12$\degree$) and (0$\degree$, -28$\degree$) in image coordinates respectively. On both of these figures there is a large un-shifted power spike at the active probing frequency (12 MHz). This is due to reflections off the window and only partially suppressed normal incidence reflections off the plasma.

No Doppler peak is visible in the spectra as obtained on single horn DBS experiments \cite{conway1,hirsch1,hirsch2,zhou1,weiwen1,tokuzawa1,pshenichnikov1,bulanin1,hillesheim2}  due to imperfect sideband suppression. In addition, SAMI cannot separate O and X-mode polarisations at present which will increase the number of different $K_{\perp}$s being sampled, thereby delocalising the scattering location. However, the directional weighting imposed by the phased array does allow a red-blue power imbalance to be measured allowing information to be attained which will be discussed further in Section~\ref{sec:pitch}. Due to a continuum of different values of $K_{\perp}$, interference between multiple backscattered signals and both O and X-mode polarisations present, a quantitative explanation of the observed spectra will require a full-wave treatment; such a study is planned using the cold-plasma full-wave code EMIT-3D \cite{williams1}.

The range of $K_{\perp}$ values which will affect SAMI spectra can be estimated using the analytic formula given in \cite{hirsch1}. For a Gaussian probing beam, the $1/e$ width of weighting functions for amplitudes in $K_{\perp}$ space is given by 

\begin{equation}
\label{eq:deltak}
\Delta K_{\perp}= \frac{2\sqrt{2}}{w}  \bigg[ 1 + \bigg(  \frac{   w^{2} k_{0}  }{   \rho  }    \bigg)^{2}  \bigg]^{\frac{1}{2}}
\end{equation}

\noindent
where $w$ is the probing beam width, $k_{0}$ is the wave vector of the probing beam in vacuum and $\rho$ is the effective curvature radius within the spot $1/\rho = 1/R_{\mathrm{plasma}} + 1/R_{\mathrm{beam}}$. The SAMI probing beam illuminates the entire vertical extent of the plasma so one can take the beam width as the height of the plasma (typically 150 cm in MAST). The typical radius of curvature for the cutoff layer in a MAST plasma is $R_{\mathrm{plasma}}\sim$1.3 m. The beam pattern for a Vivaldi antenna is approximately flat within $\pm40\degree$. So assuming that the probing antenna emits as a point source and taking a typical distance from the SAMI array to the scattering region $R_{\mathrm{beam}}\sim0.8$ m. For SAMI frequencies 10 - 35.5 GHz this gives a maximum theoretical range of $\Delta K_{\perp}$s that can effect the revived spectrum of 18.2 - 62.8 cm$^{-1}$. Note this assumes no directionality in our receiving beam which greatly increases the sensitivity to a much narrower range of $\Delta K_{\perp}$s. For example, if we assume a probing beam size as that of just the central maxima (as shown for 16 GHz in Figure~\ref{SAMIinstallation}) and taking the corresponding beam width at each frequency for a plasma 0.8 m away the range of $\Delta K_{\perp}$s reduces to 1.8-1.9 cm$^{-1}$.



\subsection{$K_{\perp}$ map}

\begin{figure}[h!]
\centering
\includegraphics[width=0.45\textwidth]{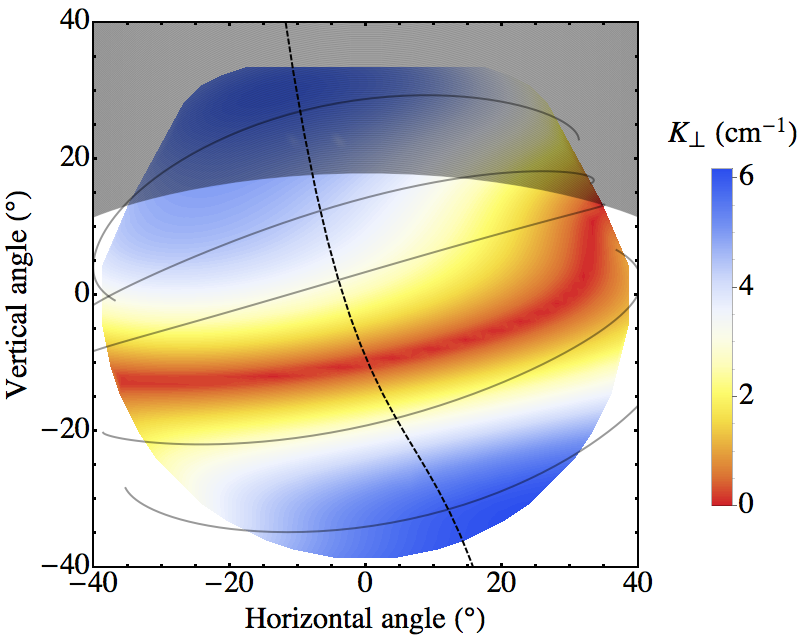}
\caption{Peak $K_{\perp}$ sensitivity for 16 GHz probing as a function of vertical and horizontal viewing angles calculated using the beam-tracing code TORBEAM 300 ms into MAST shot 27969. The diagonal continuous black lines indicate the magnetic field lines on the 16 GHz normal incidence cutoff surface. The vertical dashed black line indicates the region where probing trajectories would be perpendicular to the magnetic field.
\label{fig:k_plot}}
\end{figure}

Figure~\ref{fig:k_plot} shows the distribution of $K_{\perp}$ values as a function of probing orientation that are accessible at 16 GHz, 300 ms into MAST shot 27969 calculated using the beam-tracing code TORBEAM \cite{poli1}. Doppler backscattering is only expected when the incident beam is aligned perpendicular to the magnetic field. This is indicated in Figure~\ref{fig:reflPntPlot} by the dashed line and is consistent with observations we will consider later such as the orientation of the red-blue shifted power imbalance observed in Figure~\ref{fig:fieldlineoverplot}.

We can see from Figure~\ref{fig:k_plot} that many values of $K_{\perp}$ can be measured simultaneously using a 2D DBS although unlike other steerable monostatic DBS systems in its current configuration SAMI has been unable to measure $K$-spectra. This is a result of too few antennas in the receiving array leading to incomplete directional separation. However, the effective number of pixels in the image is approximately proportional to the square of the number of antennas. Therefore, additional antennas would result in a much reduced side-lobe level and there is no technical reason that a phased array antenna with a greater number of antennas could not make $K$-spectra measurements.

\subsection{Initial results}

\begin{figure}[h!]
\centering
\includegraphics[width=0.55\textwidth]{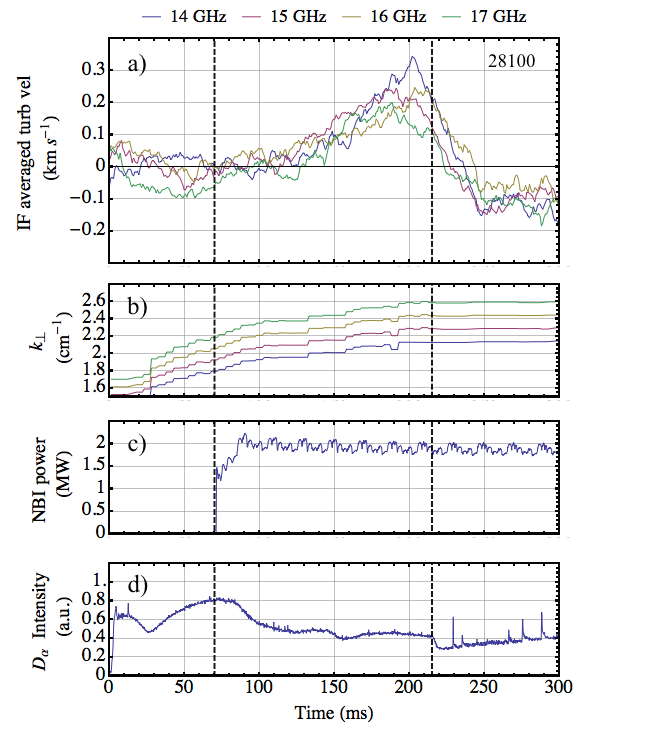}
\caption{
a) 40 ms moving average of IF averaged turbulence propagation velocity for four DBS frequency channels during MAST shot 28100. Data acquired with the receiving beam focused at (-20$\degree$, 8$\degree$) and switching in frequency every 200 \textmu s. The black vertical dashed lines at 70 and 215 ms indicate when NBI power is applied and when the plasma enters H-mode respectively. 
b) The value of the probing beam wavevector at the scattering location for a beam launched in the (-20$\degree$, 8$\degree$) direction as a function of time as calculated by TORBEAM.
c) Co-injected neutral beam power.
d) D$_{\alpha}$ emission.
\label{fig:turb_vel_fn_time}}
\end{figure}

SAMI is a proof of principle diagnostic and DBS experiments have never been attempted using a phased array previously. Limited directional weighting can result in difficult interpretation of spectra as discussed in Section~\ref{dataanalysistechniques}. However, conducting 2D DBS allows flexibility in the alignment of the probing beam. Figure~\ref{fig:turb_vel_fn_time} shows the results from a beam that was aligned with time averaged maximum blue shift during MAST shot 28100 found at -20$\degree$ on the horizontal and 8$\degree$ on the vertical. The active probing beam was at 10 MHz on the IF and data was digitised on four frequencies: 14, 15, 16 and 17 GHz with a switching time of 200 \textmu s.

Figure~\ref{fig:turb_vel_fn_time}a shows the 40 ms moving average centre of mass of the turbulence velocity across 10$\pm$0.2 MHz on the IF with the central unshifted peak (10$\pm$0.01 MHz) notched as a function of time. During these measurements the receiving beam was focused in the (-20$\degree$,8$\degree$) direction. The turbulent velocity was calculated from the observed Doppler shift using the value of $k_{\perp}$ at the scattering location as calculated by TORBEAM (Figure~\ref{fig:turb_vel_fn_time}b).

\begin{figure}[h!]
\centering
\includegraphics[width=0.525\textwidth]{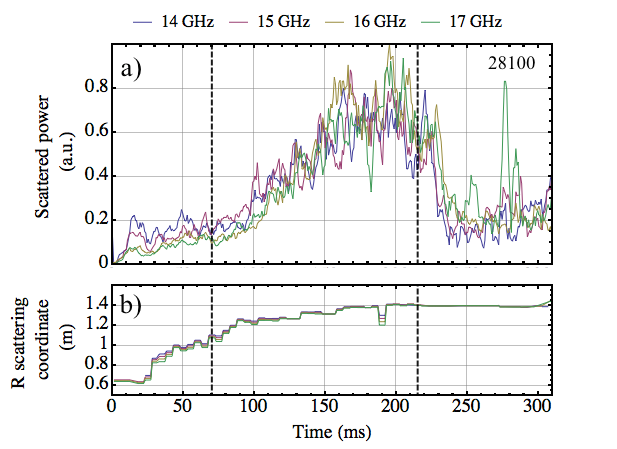}
\caption{
a) 5 ms moving average Doppler shifted power when receiving beam focused at (-20$\degree$, 8$\degree$). The two black vertical dashed lines at 70 and 215 ms indicate when NBI injection is applied and when the plasma enters H-mode respectively.
b) The major radius (R) scattering location as a function of time as calculated by TORBEAM.
\label{fig:pow_fn_time}}
\end{figure}


A gradual increase in the centre of mass turbulence velocity is observed from 70 ms onwards following 2.5 MW of NBI power being applied by one of MAST's on axis, co-injected, PINIs (Figure~\ref{fig:turb_vel_fn_time}c).  This rise in observed Doppler shift results from spin-up caused by momentum injection from the NBI system and has been observed on numerous other DBS systems, for example \cite{conway1,hillesheim2}. Once the plasma enters H-mode at 215 ms (indicated by a decrease in D$_{\alpha}$, Figure~\ref{fig:turb_vel_fn_time}d) there is an abrupt change in the sign of the observed turbulence velocity. A sharp change in the turbulent velocity coinciding with the onset of H-mode has also been measured on previous DBS experiments \cite{hillesheim2,conway1} and results from the edge turbulence velocity being dominated by toroidal fluid velocity during L-mode and diamagnetic velocity in H-mode. The steep pressure gradient that forms in the edge region during H-mode results in a sharp increase of the diamagnetic velocity \cite{conway1}.

Figure~\ref{fig:pow_fn_time}a shows the 5 ms moving average of the total Doppler shifted power. The power plotted is the summed power across the IF 10$\pm$0.2 MHz where the central unshifted frequency peak is notched (10$\pm$0.01 MHz) and the average background noise level is subtracted. The Doppler power steadily increases after the NBI is applied. This is likely to be caused by an increase in the density and the scattering location moving closer to the SAMI array (Figure~\ref{fig:pow_fn_time}b) and/or an increase in the turbulence amplitude. This is followed by a sharp drop in power as the plasma enters H-mode despite the scattering location not changing significantly (Figure~\ref{fig:pow_fn_time}b). This decrease is caused by the suppression of turbulence in the edge region. The ramp up in power during NBI injection and drop in power as the plasma enters H-mode has been observed in other DBS experiments \cite{hillesheim1}.  In Figure~\ref{fig:pow_fn_time}a microwave bursts are observed after the plasma enters H-mode as each Edge Localised Mode (ELM) coincides with microwave emission up to four orders of magnitude above thermal \cite{freethy2}. The average background emission is subtracted but this will only nullify the ELM emission if the burst is evenly distributed across the IF.

SAMI has observed trends in turbulence velocity and DBS power which have been observed on previous experiments as demonstrated in  Figures~\ref{fig:turb_vel_fn_time} and \ref{fig:pow_fn_time}. Though SAMI has some notable limitations in its current form, the results presented here are encouraging for the future feasibility of 2D DBS systems.

\subsection{Magnetic pitch angle measurements}
\label{sec:pitch}

2D DBS experiments using a phased array could potentially provide a way of measuring the magnetic pitch angle profile with high temporal and spatial resolution. 

\begin{figure}[h!]
\centering
\includegraphics[width=0.5\textwidth]{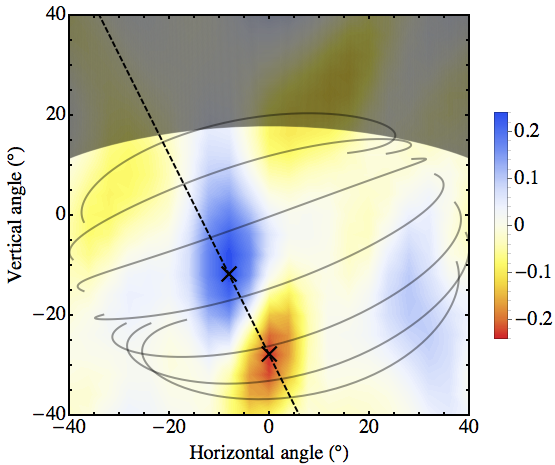}
\caption{
Doppler shifted power difference at 16 GHz during MAST L-mode shot 27969 at 300 ms. Blue and red indicate more blue than red and more red than blue shifted power respectively. Magnetic field lines on the corresponding normal incidence O-mode cutoff are over-plotted in grey. The maxima and minima in the power deficit between red and blue shifted power are marked with black crosses. The dashed black straight line connects the blue and red maxima. Magnetic field line information and the location of the normal incidence O-mode cutoff surfaces are provided by EFIT and Thomson Scattering respectively.\label{fig:fieldlineoverplot}}
\end{figure}

To form Figure~\ref{fig:fieldlineoverplot} the SAMI array was focused, using beam forming, onto each point in an equally spaced 21 by 21 grid spanning $\pm$40$\degree$ in the horizontal and vertical viewing directions. This was done using data acquired 300 ms into MAST shot 27969 whist actively probing the plasma at 16 GHz. For each grid point the spectra of the received beam is analysed. In Figure~\ref{fig:fieldlineoverplot} the difference between the blue and red shifted power is plotted after the background passive emission has been subtracted and the central unshifted peak (10 $\pm$ 0.01 MHz) has been notched. Net positive and negative regions show where more blue and red shifted power is present respectively. The colour bar is normalised to the total Doppler shifted power received so that 0.2 indicates that the surplus of blue shifted power at that point represents 20$\%$ of the total Doppler shifted power received. The two regions which show the most red-blue power imbalance are marked with black crosses; the spectra observed at these locations are shown in Figures~\ref{fig:spectra}a and \ref{fig:spectra}b. The black dashed straight line connects the points of maximum blue and red-shifted power imbalance. The magnetic field lines, as calculated from EFIT and Thomson scattering, are over-plotted in grey. 

One would expect Bragg backscattering to occur only when the probing beam is perpendicular to the magnetic field lines. Therefore, the orientation of the red and blue maxima allow a pitch angle measurement to be made. SAMI is the first DBS system to make a pitch angle measurement. This new capability is a direct result of simultaneous 2D imaging made possible by the use of a phased array.

\begin{figure}[h!]
\centering
\includegraphics[width=0.5\textwidth]{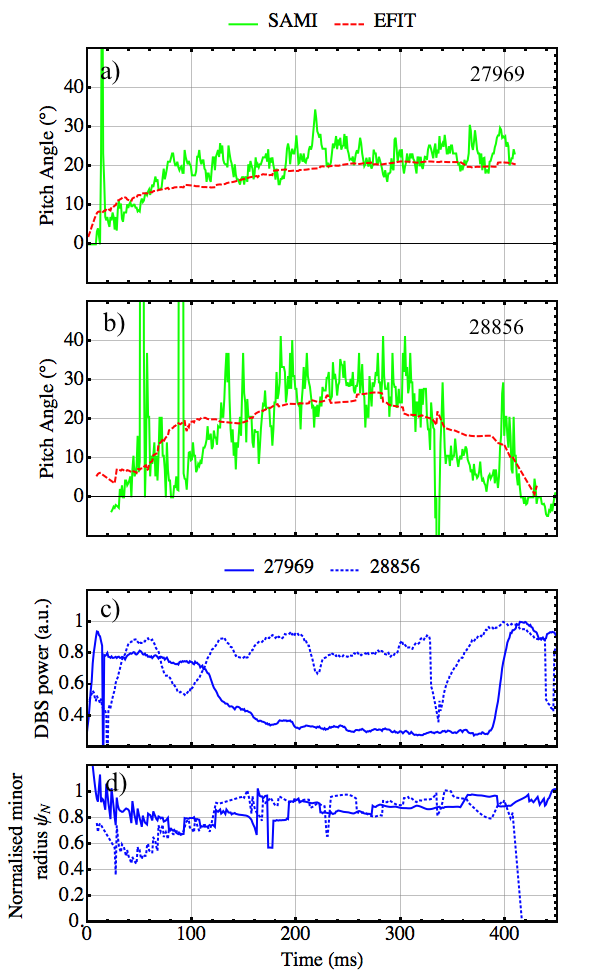}
\caption{
Magnetic pitch angle as measured by SAMI and EFIT at 16 GHz and 10  GHz during MAST shots 27969 a) and 28856 b) respectively. The SAMI pitch angle measurement is  shown by the solid green line. The magnetic pitch angle as calculated by EFIT is shown by the dashed red line. Each SAMI pitch angle measurement was calculated using 8 ms of data. The EFIT pitch angle was evaluated at the scattering location of a ray launched directly in-between the locations of the red and blue-shifted maxima and minima at each moment in time as calculated by TORBEAM. c) Shows the backscattered power level at the red and blue-shifted Doppler power maxima during shot 27969 (continuous blue line) and 28856 (dashed blue line). Note that the power levels for each shot use different normalisations. d) The normalised minor radius ($\psi_{\mathrm{N}}$) of the scattering location for shots 28856 and 27969 as calculated by TORBEAM.
} 
\label{fig:27969_pitch}
\end{figure}

Figures~\ref{fig:27969_pitch}a and b show the magnetic pitch angle as measured by SAMI (green line) and EFIT (dashed red line) as a function of time for MAST shot numbers 27960 (fixed frequency 16 GHz) and 28856 (fixed frequency 10 GHz). The SAMI pitch angle time evolution was calculated using an 8 ms sliding data window. The EFIT pitch angle was calculated at the scattering location of a ray launched along a path directly in-between the locations of maxima and minima in power deficit at each moment in time as calculated by TORBEAM. Large discrepancies in the agreement between the SAMI and EFIT pitch angle measurements are seen to occur when the backscattered power level is low (Figure~\ref{fig:27969_pitch}c). It is worth noting that the optimum configuration for SAMI to make pitch angle measurements was not known when the data for these shots was taken. Nevertheless, despite being a proof of principle, first of its kind diagnostic, SAMI has made pitch angle measurements which agree well with EFIT considering SAMI's present limitations. The amount of uncertainty in the SAMI and EFIT pitch measurements is left the subject of a future publication. The evolutionary trend in the pitch, as measured by SAMI and EFIT, is in agreement throughout both shots. The fluctuation level in the SAMI pitch angle measurements during shot 28856 is noticeably higher than during 27969. This is expected as being a low frequency 10 GHz shot the plasma frequency cutoff during 28856 was further out in the scrape off layer where the density fluctuation level is greater resulting in a noisier measurement. Though the radial scattering locations are similar in both shots (Figure~\ref{fig:27969_pitch}d) the pitch evolution varies due to different plasma current temporal profiles. It is also apparent from Figure~\ref{fig:27969_pitch}d that scattering takes place in the edge region consistent with Figure~\ref{fig:reflPntPlot}.

As will be discussed further in Section~\ref{sec:furtherwork} there are numerous ways that SAMI can be upgraded and reconfigured to improve the accuracy of the pitch angle measurements.

\section{Conclusion}
\label{sec:conclusion}

SAMI has been used to conduct the first ever simultaneous 2D DBS experiments and has explored the feasibility of using a phased array to conduct DBS on fusion plasmas. SAMI has measured phenomena that have been predicted by theory and observed on previous conventional DBS experiments. An increase in the observed turbulence velocity with application of NBI and a sharp transition in velocity during the L-H transition (see Figure~\ref{fig:turb_vel_fn_time}a) have been observed on MAST. Doppler shifted power has been seen to increase after application of NBI and drops when the plasma enters H-mode (see Figure~\ref{fig:pow_fn_time}a). As well as reaffirming measurements made by conventional DBS systems, conducting 2D Doppler experiments has allowed the magnetic pitch angle to be measured, a parameter which has not been probed by any previous DBS system. 

Conventional DBS systems have to be aligned at a specific orientation so that their probing beams are perpendicular to the magnetic field at the scattering location. These systems are limited as this orientation will only be optimal for one particular magnetic pitch angle. This problem is exacerbated on spherical tokamaks where the pitch angle may vary considerably (see Figures \ref{fig:27969_pitch}a, \ref{fig:27969_pitch}b). As SAMI is always probing in every direction this problem is bypassed and the spatial variability of the backscattering maxima allows a magnetic pitch angle measurement to be made.  

Therefore SAMI has shown that it is feasible to conduct DBS experiments using a phased array and that using a 2D system not only allows flexibility in the directionality of the beam, but also allows new parameters to be measured. The main limitations of the current SAMI system are that it has only eight antennas and cannot separate O and X-mode polarisations.


\section{Further work and discussion}
\label{sec:furtherwork}

SAMI is a prototype 2D DBS system and after the initial results presented in this paper it is apparent that there are many options for further development. 

An improvement to the SAMI system would be to increase the number of receiving antennas. The effective number of pixels in the image is proportional to the square of the number of antennas. Therefore extra antennas would greatly improve the directional weighting that SAMI could apply and therefore increase the accuracy of turbulent velocity measurements as stray backscattered radiation and velocities outside the focused beam would be further suppressed. Accurate measurements of toroidal rotation velocity could then be compared with Charge eXchange Recombination Spectroscopy (CXRS) and Beam Emission Spectroscopy (BES) diagnostics. SAMI is currently unable to measure $K$-spectra as measuring the amplitude of a particular $K$ requires probing the plasma at a single location. Extra antennas would result in better probing beam localisation making $K$-spectra measurements possible. Exactly how many more antennas are required in order to make accurate $K$-spectra measurements requires an involved treatment of antenna optimisation and has many variable factors such as amplitude of the turbulence and plasma geometry and is therefore left as the subject of a future publication. 

The arrangement of the antennas in the receiving array could also be changed so that the array was optimised for conducting DBS experiments. For example, the receiving array could be arranged linearly with the antennas aligned with the magnetic pitch angle in order to attain improved spatial directionality along the axis of the array which would allow for increased $K_{\perp}$ selectivity. The limitation of this technique is that the array would only be optimised for a particular pitch angle and, as is evident in Figures~\ref{fig:27969_pitch}a and \ref{fig:27969_pitch}b, on spherical tokamaks the magnetic pitch angle is highly variable. 

The SAMI system could also be improved by enabling polarisation separation. All data presented here was obtained using linearly polarised Vivaldi antennas with one polarisation orientation. This means that O and X-mode radiation cannot be separated. It can be seen in Figure~\ref{fig:reflPntPlot} that the O-mode and X-mode cutoffs are in close proximity during SAMI experiments conducted on MAST; therefore interference effects might be significantly affecting the backscattered signal. This is a potential source of error in the turbulent velocity measurements (Figure~\ref{fig:turb_vel_fn_time}a) and pitch angle measurements (Figure~\ref{fig:27969_pitch}). In order to attain polarisation separation it is planned to upgrade SAMI's receiving array to dual polarised PCB sinuous antennas. Polarisation separation will then be achieved by fast switching between the two orthogonal polarisations.

When the existing SAMI data set was acquired it was not known how the SAMI system would be optimised for DBS experiments. In Figures~\ref{fig:27969_pitch}a and \ref{fig:27969_pitch}b pitch angle at one frequency only is plotted. This results from the switching time between frequencies being set too short (10 - 250 \textmu s). Therefore time integration was not long enough for a pitch angle measurement to be made during switching frequency data acquisitions. Now it is known that longer time integration is required ($\sim$10 ms), experiments can be conducted providing pitch angle profiles as a function of time using the existing SAMI system. Effects on pitch angle by NBI and H-mode can then be investigated along with a comparison against Motional Stark Effect (MSE) pitch profiles once a larger, multi-frequency, data set is attained. This data will be acquired once SAMI is installed on NSTX-U in the summer of 2015.

\section{Acknowledgements}

This work was funded, in part, by EPSRC under grants EP/H016732 and EP/K504178, the University of York, and the RCUK Energy Programme under grant EP/I501045. This work has been carried out within the framework of the EUROfusion Consortium and has received funding from the Euratom research and training programme 2014-2018 under Eurofusion project ER-WP15\_CCFE-03. The views and opinions expressed herein do not necessarily reflect those of the European Commission. To obtain further information on the data and models underlying this paper please contact PublicationsManager@ccfe.ac.uk.


\section{References}
\nocite{*}
\bibliography{bibliography}
\bibliographystyle{unsrt}

\end{document}